# Second harmonic generation in germanium quantum wells for nonlinear silicon photonics

Jacopo Frigerio[1], Chiara Ciano[2], Joel Kuttruff[3], Andrea Mancini[4,5], Andrea Ballabio[1], Daniel Chrastina[1], Virginia Falcone[1], Monica De Seta[2], Leonetta Baldassarre[5], Jonas Allerbeck[3,6], Daniele Brida[6], Lunjie Zeng[7], Eva Olsson[7], Michele Virgilio[8*] and Michele Ortolani[5*]

*1* L-NESS, Dipartimento di Fisica, Politecnico di Milano, Polo di Como, Via Anzani 42, I-22100 Como, Italy

2 Dipartimento di Scienze, Università di Roma Tre, Viale Marconi 446, I-00146 Rome, Italy.

3 Department of Physics and Center for Applied Photonics, University of Konstanz, D-78457 Konstanz, Germany.

4 Nanoinstitute Munich, Ludwig-Maximilians-Universität Munich, Königinstrasse 10, 80539 Munich, Germany

5 Dipartimento di Fisica, Sapienza Università di Roma, Piazzale Aldo Moro 5, I-00185 Rome, Italy

6 Faculté des Sciences, de la Technologie et de la Communication, Université de Luxembourg, L-1511 Luxembourg

7 Department of Physics, Chalmers University of Technology, Fysikgränd 3, 412 96 Gothenburg, Sweden

8 Dipartimento di Fisica "E. Fermi", Università di Pisa, Largo Pontecorvo 3, I-56127 Pisa, Italy

[*michele.virgilio@unipi.it, michele.ortolani@roma1.infn.it](mailto:)

**Abstract.** Second-harmonic generation (SHG) is a direct measure of the strength of second-order nonlinear optical effects, which also include frequency mixing and parametric oscillations. Natural and artificial materials with broken center-of-inversion symmetry in their unit cell display high SHG efficiency, however the silicon-foundry compatible group-IV semiconductors (Si, Ge) are centrosymmetric, thereby preventing full integration of second-order nonlinearity in silicon photonics platforms. Here we demonstrate strong SHG in Ge-rich quantum wells grown on Si wafers. The symmetry breaking is artificially realized with a pair of asymmetric coupled quantum wells (ACQW), in which three of the quantum-confined states are equidistant in energy, resulting in a double resonance for SHG. Laser spectroscopy experiments demonstrate a giant second-order nonlinearity at mid-infrared pump wavelengths between 9 and 12 μm. Leveraging on the strong intersubband dipoles, the nonlinear susceptibility $\chi^{(2)}$ almost reaches $10^5$ pm/V, four orders of magnitude larger than bulk nonlinear materials for which, by the Miller's rule, the range of 10 pm/V is the norm.

**Introduction**

In recent years many research efforts have been devoted to the study and exploitation of nonlinear optical phenomena in silicon photonics integrated circuits (PICs) at infrared (IR) frequencies *[1, 2]* for applications in all-optical signal processing *[3]*, spectroscopy *[4]* and quantum optics *[5, 6]*. In this context, second-order nonlinear optics is essential for many classical and quantum applications, from high-speed optical modulation via the Pockels effect *[7]*, to *f*-2*f* frequency-comb self-referencing *[8]*, and even to direct frequency-comb generation by cascaded SHG events that can mimic third-order nonlinear effects *[9]*. Bulk Si and Ge are compatible with PIC foundry processes *[10, 11]*, however they feature vanishing second-order susceptibility $\chi^{(2)}$ due to their centrosymmetric unit cell. Several approaches have then been proposed to achieve second-order nonlinearities in silicon PICs: Electric field bias can induce nonlinearity in periodically poled silicon-on-insulator (SOI) waveguides, however with a rather small $\chi^{(2)} = 0.64$ pm/V *[12]*; SHG has been observed in $Si_3N_4$ waveguides on $SiO_x$ featuring $\chi^{(2)} = 2.5$ pm/V *[13]*, and more recently in micro-resonators *[14]*. The transparency range of both SOI and $Si_3N_4$ waveguides is however limited to wavelengths $\lambda < 4$ μm as a result of the strong phonon absorption of $SiO_x$ *[15, 16, 17]*. Proposed unconventional approaches to generate optical nonlinearity in bulk Si include nanostructures with strain gradients *[18]*, which are difficult to realize in a controlled manner, and nonlinear free-electron plasma oscillations in heavily doped Si or Ge *[19]*, which are accompanied by high ohmic losses. A different solution for second-order nonlinearity in Si PICs is here provided by asymmetric-coupled quantum wells (ACQWs) made of SiGe epitaxial layers grown on Si substrates (Fig. 1(a)). ACQWs are a type of semiconductor heterostructure based on two quantum wells of different thickness and/or composition, separated by a thin tunneling barrier (Fig. 1(b)). Electrons or holes confined into the ACQW planes (the epitaxial growth direction being normal to those planes) populate quantized levels arranged in discrete subbands, i.e. "copies" of either the valence or the conduction band separated by the differences in quantization energy $E_i$ (Fig. 1(c)). In ACQWs, either holes or electrons provide the dipole strength for the nonlinear interaction of the material with the laser pump through their inter-subband transitions (ISBTs) [20, 21]. The ACQW subband structure can be specifically designed for resonant enhancement of a specific nonlinear effect (here SHG) and they can behave as artificial nonlinear materials with high efficiency [22, 23, 24]. It is important to notice that the ACQW second-order nonlinearity is purely based on the electron wavefunctions being asymmetrically delocalized in the two coupled wells, and not on the crystal lattice asymmetry as in natural nonlinear materials. Then, the Miller's rule, a phenomenological argument for estimating nonlinear susceptibilities based on the value of the crystal lattice parameter, does not hold for ACQWs [20, 21], releasing the physical limit setting the value of $\chi^{(2)}$ in the range of 10 pm/V for all existing natural nonlinear crystals [5]. One can then quantum-design the ACQW wavefunctions and obtain a giant $\chi^{(2)}$ for SHG up to $10^5$ pm/V [22, 23, 24, 25], exploiting the key fact that, in an ideal ISBT, the joint optical density of states is a Dirac delta function, because all initial and final subband states have exactly the same energy distance, unlike in a conventional interband transition. The design criterion for SHG in ACQWs is that the ISBT energies $h\nu_{12}$ and $h\nu_{23}$ (where $h\nu_{ij}$ is the energy of the ISBT between $i$ and $j$, hereafter indicated as $i \rightarrow j$) should both be in resonance with the pump photon energy $h\nu_{pump}$, or $h\nu_{12} \cong h\nu_{23} \cong h\nu_{pump}$. Setting the $z$ axis parallel to the carrier confinement direction (i.e. orthogonal to the QW growth plane), the $\chi^{(2)}_{zzz}(2\nu)$ element of the second-order susceptibility tensor can be conveniently approximated, assuming parabolic subband dispersion, by the following expression [20]:

$$\chi^{(2)}_{zzz}(2\nu) = \frac{q^3(N_1 - N_2)}{\epsilon_0 h^2} \frac{z_{12} z_{23} z_{31}}{(\nu - \nu_{12} - i\gamma)(2\nu - \nu_{13} - i\gamma)} \quad (1)$$

where $N_i$ is the 3D population density of subband $i$, $z_{ij}$ are the vertically oriented dipole moments of $i \rightarrow j$ and $\gamma$ is the intrinsic linewidth assumed identical for all ISBTs. The numerator $z_{12} z_{23} z_{31}$ is nonzero only in the presence of an asymmetry of the QW potential profile that breaks the ISBT selection rule $i \rightarrow 2ni+1$, as in ACQWs [20,21]. In the ideal case of parabolic subband dispersion, $\nu_{ij}$ does not depend on the in-plane crystal momentum $k_{//}$ and the SHG doubly-resonant condition is obtained at the same $\nu_{pump}$ for all electrons (or holes) populating the involved subbands, while in real cases one has to integrate over $k_{//}$ in the entire 2D Brillouin zone and the SHG resonance spectrum can broaden considerably [25]. In summary, not only the ACQW structure breaks the inversion symmetry as required for SHG, but it also provides the freedom of wavefunction design to achieve doubly resonant second-order electromagnetic interaction [21]. In addition, the doping level sets the strength of the nonlinearity. The doping can be introduced by chemical substitution (in this work B for Ge) or it can be modulated at the picosecond time scale with a near-IR optical pump [26].

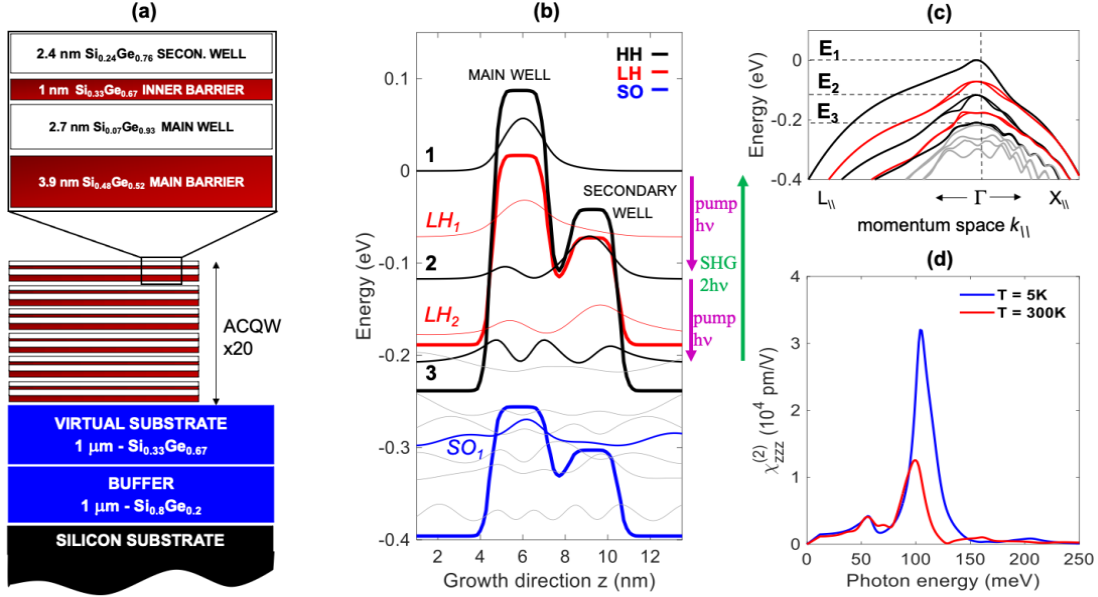

*Figure 1*. (a) Schematic of sample A showing the different epitaxial block thickness and composition: buffer, virtual substrate, ACQW stack repeated 20 times. (b) Valence band-edge profiles of sample A forming the main and secondary wells (thick lines, HH: heavy holes; LH: light holes; SO: split-off band), the subband wavefunctions (thin lines) and the SHG transitions (arrows; for holes, wells correspond to band-edge potential maxima and transitions are downwards in energy). (c) In-plane dispersion of the hole states in the first Brillouin zone around the $\Gamma$ point as a function of the in-plane wavevector $k_{||}$. (d) Calculated second-order nonlinear susceptibility $\chi^{(2)}_{zzz}$ as a function of the pump photon energy.

SHG has been systematically observed in ACQWs *[20, 22, 23]* made of III-V compound semiconductor materials, and, in a pioneering work, also in Si-rich SiGe ACQWs *[27]*. Recently, SHG efficiency enhancements of three orders of magnitude have been reported in III-V compound ACQWs using plasmonic nanoantennas [*23, 28*] and diffraction grating couplers [*24*] for electric field enhancement, however the integration of ACQW nonlinearity into silicon PICs has never been targeted before. In this work, we demonstrate SHG in Ge-rich SiGe ACQWs grown on Si substrates (as shown in Fig. 1(a)) using silicon-foundry-compatible processes. The epitaxial deposition of Ge-rich SiGe heterostructures on silicon has undergone an impressive development in the last decade driven by the envisioned applications in telecom and datacom. Optical modulators based on Ge/SiGe multiple quantum wells have been reported *[29, 30, 31]* and Ge/SiGe quantum cascade lasers are under investigation *[32, 33]*. In the case of nonlinear PICs, the use of Ge-rich heterostructures will be necessary to realize integrated rib or ridge waveguides, which is a requirement to obtain long interaction regions for efficient SHG in PICs without having to use very high pump power densities. Indeed, Ge/SiGe epitaxial layers forming the ACQWs (refractive index *n* close to 4.0) can be etched to form the waveguide core on top of the Si substrate ($n = 3.4$), operating as the waveguide cladding together with the air or the dielectric environment, an ideal configuration for on-chip mid-IR spectroscopy applications *[34, 35, 36, 37]*.

**Wavefunction design.**

The interband transition edge of Ge in the near-IR at 0.66 eV together with its high transparency region extending to $\lambda < 15$ μm, make it the ideal material for mid-IR PICs, whereas intrinsic absorption losses limit waveguiding in pure Si to $\lambda < 8$ μm *[38, 39, 40]*. Here, we aimed at demonstrating the nonlinear Ge/SiGe ACQW operation in a broad mid-IR wavelength range: laser pump wavelengths were selected in the 12 to 9.2 μm range (SHG wavelengths from 6.0 down to 4.6 μm). We leverage on hole-doped rather than electron-doped structures in order to exploit the larger valence band offsets up to 0.5 eV. It is worth to notice that ISBTs in hole-doped materials, at odds with more common electron-doped ACQWs, also feature non-vanishing in-plane components of the optical transition dipoles connecting the heavy hole (HH), light hole (LH) and split-off (SO) valence bands. Consequently, one could achieve also large off-diagonal tensor components $\chi^{(2)}_{xzk}(2\nu)$ thus opening interesting opportunities to realize polarization mixing in nonlinear phenomena [25, 41]. In this work, however, we have designed and investigated samples with an epitaxial structure specifically optimized for SHG featuring three energy levels $E_1$, $E_2$, $E_3$ belonging to the HH band (**well**/barrier thickness **2.4**/1.0/**2.7**/3.9 nm, and Ge content $x_{Ge}$ = **0.76**/0.67/**0.93**/0.52, see Fig. 1). The levels are equally spaced: $h\nu_{12} = h\nu_{23}$. Sample design parameters are reported in Table 1: if compared to the ideal case of sample A, sample B has half the doping level, sample C has its $E_3$ purposely offset from double SHG resonance ($h\nu_{12} \neq h\nu_{23}$), and sample D is undoped. Samples B, C, D are used for comparison to define the performance of the ideal sample A, whose predicted $\chi^{(2)}_{zzz}(2\nu)$ is reported in Fig. 1(d) (other tensor components $\chi^{(2)}_{zjk}(2\nu)$ are reported in Supplementary Information S1).

**Table 1**. Summary of design and measured parameters of the four samples

| Sample | Wavefunction design | Main well thickness $t_w$ (nm) | Acceptor density $N_a$ (cm$^{-2}$) | Measured hole density $N_{tot}$ (cm$^{-2}$) | Measured SHG efficiency $K_2$ (1/W) |
|---|---|---|---|---|---|
| **A** | Doubly resonant | 2.7 | 4.2 ×10$^{11}$ | (3.9±0.3) ×10$^{11}$ | 5.7×10$^{-3}$ |
| **B** | Doubly resonant | 2.7 | 2.1×10$^{11}$ | (2.5±0.3) ×10$^{11}$ | 1.4×10$^{-3}$ |
| **C** | Single resonance | 3.0 | 4.2 ×10$^{11}$ | (3.9±0.3) ×10$^{11}$ | 0.6×10$^{-3}$ |
| **D** | Reference | 2.7 | undoped | - | < 0.1×10$^{-3}$ |

**Heterostructure growth and characterization**

The samples have been grown by low-energy plasma enhanced chemical vapor deposition (LEPE-CVD) *[42]* on Si(001) wafers, a standard for PICs (see Supplementary Information S2). The superlattice consists of 20 periods of the ACQW stack shown in Fig. 1(a). Such low number of ACQW repetitions has been chosen to optimize the linear dichroism effect of ISBTs in IR transmission spectroscopy *[33]*, but it should be pointed out that the growth procedure can be straightforwardly extended to deposit hundreds of ACQW periods *[43]* required to fill the ridge height > 2 µm of a mid-IR integrated waveguide *[38]*. A Si-rich buffer layer is deposited at the interface with the Si substrate to reduce the lattice mismatch (hence the threading dislocation formation at the initial stage of the relaxation) *[44]*. A Ge-rich layer is then deposited to serve as a virtual substrate for the ACQWs. High resolution X-Ray diffraction (HR-XRD) has been used to measure the Ge content, the strain and the in-plane lattice parameters for all samples. A clear superlattice period is observed indicating regular periodicity of the ACQW structure (Fig. 2(a)-(b), see Methods). The scanning transmission electron microscopy (STEM) image in Fig 2(c), (e) shows ACQWs with sharp interfaces arranged in identical subsequent periods along the growth direction *z*. The corresponding energy-dispersive X-Ray emission spectroscopy (EDS) data (red marks in Fig. 2(f)) confirm the different Ge content in the two wells obtained by fitting the HR-XRD data (blue curve in Fig. 2(f)), which is crucial to break the inversion symmetry for SHG.

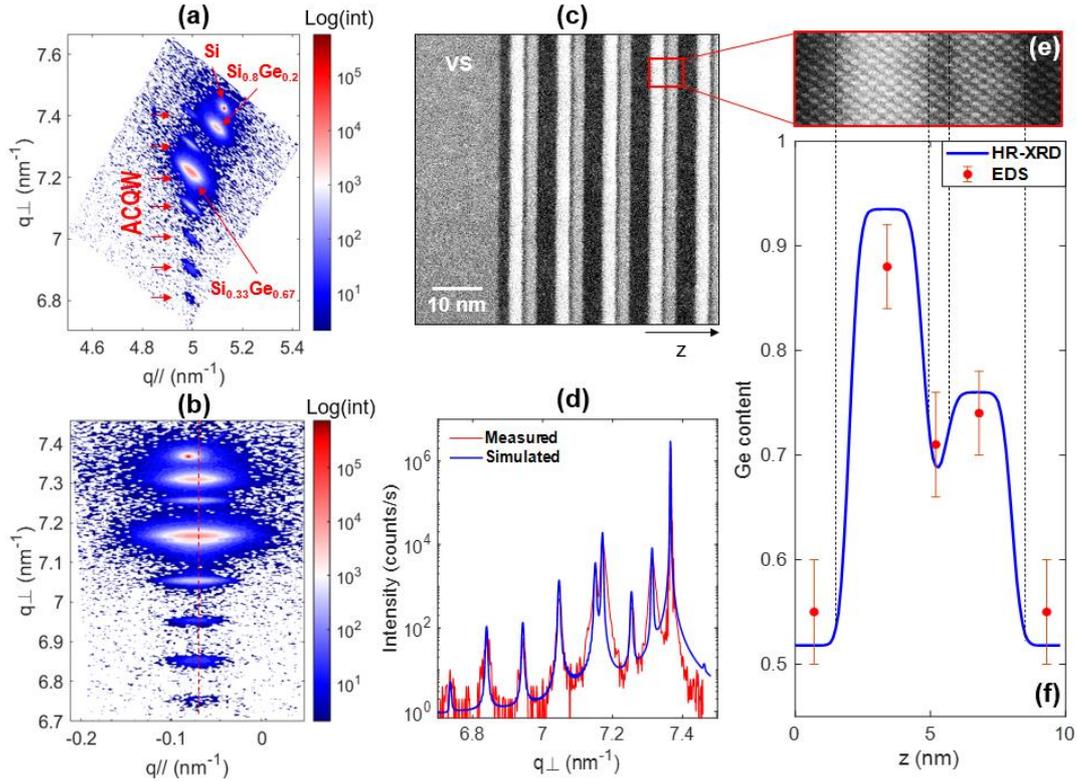

*Figure 2.* (a, b) HR-XRD reciprocal space maps of sample A with respect to the (224) and (004) Si reflections. (c) STEM images of a different piece of sample A. (d) ω-2θ scan of the Si(004) reflection (red) together with Darwin model simulations (blue) performed to extract the Ge composition of each layer and the intermixing depth. (e) High-resolution STEM, greyscale intensity is proportional to local Ge content. (f) Ge content profile of sample A retrieved from HR-XRD measurements (blue curve) and by EDS measurements (red dots).

Ge/SiGe heterostructures with narrow wells, abrupt interfaces and high compositional mismatches between adjacent layers pose significant challenges in terms of epitaxial deposition. To this aim one has to consider that, unlike III-V semiconductors, Si and Ge are completely miscible over the entire compositional range. As a consequence, the realization of atomically sharp interfaces is hindered by entropic intermixing of Si into pure Ge layers, and Ge into SiGe layers. Therefore, the pure Ge/SiGe heterostructures proposed in Ref. [*25*] could not be grown, because the actual compositional profile is smoothed out by intermixing and, in the case of the main well, also by the presence of residual SiH4 gas in the CVD reactor. The resulting ACQW potential profile of sample A shown in Fig. 1(b) represents the best possible compromise, with $h\nu_{12} = h\nu_{23} = 105$ meV and a doubly-resonant giant $\chi^{(2)}_{zzz}$ up to $3\times10^4$ pm/V as seen in the calculation of Fig. 1(d), in which the subband energy separation dependence on $k_{//}$ has been taken into account in order to compute $\chi^{(2)}_{zzz}(2\nu)$ by integrating over the crystal momentum in the entire 2D Brillouin zone. Although the pure Ge/SiGe heterostructure design proposed in Ref. [*25*] had $h\nu_{12} = h\nu_{23} = 120$ meV and $\chi^{(2)}_{zzz}$ up to $1\times10^5$ pm/V [*25*], the presently expected values of $\chi^{(2)}_{zzz}$ are still three orders of magnitude higher than those of typical nonlinear crystals.

**Linear spectroscopy and pump wavelength selection**

The linear absorption by ISBTs in the doped samples A, B, C was measured by Fourier Transform Infrared spectroscopy (FTIR) in linear dichroism mode to filter out the substrate contributions, with the samples prepared in the prism-shaped slab-waveguide configuration as sketched in Fig. 3(a), where the TM polarization senses the ISBTs with dipole along *z* and the TE polarization is used for internal reference. In the dichroic absorption spectra of Fig. 3(b) one can observe that all samples display two ISBT peaks of similar intensity: the low frequency peak around 32 THz is naturally assigned to 1→2 and the high frequency peak (58 THz in samples A, B and 53 THz in C) to 1→3. Note that sample C was intentionally detuned by design from the double resonance condition, but also in samples A and B the perfect double resonance condition $h\nu_{13} = 2h\nu_{12}$ is only approximately realized. The best-fit Lorentzian frequencies $\nu_{ij}$, oscillator strengths $f_{ij}$, and half-linewidths $\gamma_{ij}$ of the absorption peaks are reported in Supplementary Information S3.

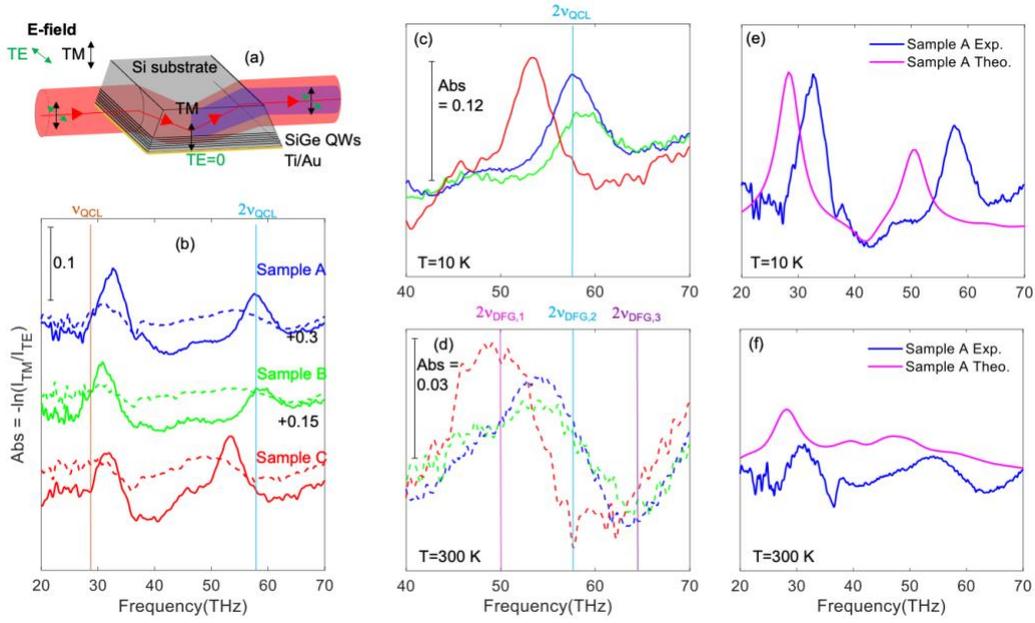

*Figure 3. Linear dichroic absorption spectroscopy in the mid-IR. (a) Sketch of the surface-plasmon waveguide structure with prism-shaped slab geometry. The electric field in the ACQW region is mostly vertical (TM mode), while the TE mode transmittance serves as spectral reference. (b) Dichroic absorption -ln(TM/TE) at T = 10 K (continuous lines) and T=300 K (dotted lines). (c, d) zoom on the ISBT peak at $\nu_{23}$ with indication of the different laser pump frequencies used in this work. QCL: quantum cascade laser, used with the samples kept at 10 K. DFG: difference-frequency generation, used with the samples kept at 300 K. (e,f): Direct comparison of the calculated (magenta curve) and measured (blue curve) dichroic absorption spectra. The vertical scale and offset have been adjusted for each curve.*

The main effect of heating the samples to 300 K is the notable broadening of the ISBT linewidths as shown in Fig. 3(d). At room *T*, due to thermal excitation, carriers are spread in *k* space around the Γ point, hence the non-parabolicity of the subbands produces a non-trivial dependence of the $\nu_{ij}$ values on $k_{//}$, which is the main factor contributing to the broadening of the linewidths. This effect is also responsible for the temperature-induced redshift of the 1→3 absorption peak. In fact, the population of the fundamental subband at larger $k_{//}$ due to thermal excitation activates different 1→3 transitions featuring lower photon energies, as it can be argued from the non-parabolicity of the band structure shown in Fig. 1(c). Intrinsic line broadening effects at finite-$T$ are expected to constitute the main limitation to the SHG power that can be obtained from doubly resonant ACQWs.

In Fig. 3(c) and 3(d) the 1→3 ISBT spectra are re-plotted with indication of the selected pump photon frequencies for the experiments described below employing a quantum cascade laser (QCL) pump at 10 K and a difference-frequency generation (DFG) pump at 300 K. One can identify in Fig. 3(c) the perfect resonance of $2\nu_{QCL}$ = 58.6 THz with $\nu_{13}$ of samples A and B. Lastly, we compare in Figs. 3(e) and 3(f) the theoretical and experimental dichroic absorption spectra for sample A. The overall spectral shape and the absorption intensities are very well reproduced by our calculations at both $T$ values. The dip at 40 THz is a signature of the TE-polarized ISBT 1→LH$_2$ in good agreement with the energy difference of 175 meV observed in Fig.1(a). The rigid frequency offset of approximately 7 THz for the two TM-polarized peaks 1→2 and 1→3 can instead be attributed to many-body depolarization effect, not included in our calculations (see Methods), which induces a blueshift of the TM-polarized absorption frequencies if compared to the bare bandstructure energy differences in Fig. 1(a). Note that the values to be inserted in Eq. 1 are the actual absorption frequencies, not the bare energy differences.

**Measurement of absolute SHG efficiency**

The demonstration of the doubly resonant nonlinearity in Ge/SiGe ACQWs is here provided by experimental comparison of samples A and C at 10 K with different $\nu_{13}$ but otherwise identical relevant parameters: $\nu_{12}$, doping, and number of periods. We used $\nu_{QCL}$=29.3 THz ≈ $\nu_{13}/2$ for sample A, however there is a slight detuning $\delta=\nu_{12}-\nu_{QCL}$ = 3 THz ≈ $\gamma_{12}$ that reduces the pump intensity depletion while travelling through the waveguide: this choice of $\nu_{QCL}$ maximizes the SHG efficiency difference between samples A and C and also the total SHG intensity, as we shall see below.

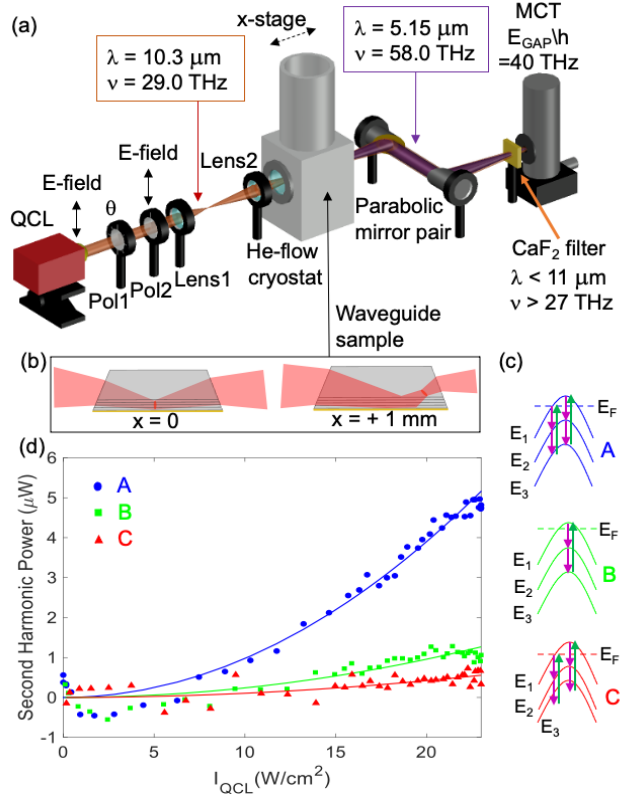

*Figure 4. (a) Scheme of the nonlinear experiment for absolute characterization of SHG efficiency. In the TM mode, the radiation electric field is oriented along the ISBT dipole, while the TE mode intensity serves as reference spectrum. (b) The SHG emission disappears when the sample is shifted by +/-1 mm (twice the estimated depth of focus) because the electric field value drops quickly when the ACQWs are out of focus. This technique was used to subtract the contribution of the residual pump photons to the detector signal. (c) Simplified sketch of the SHG process in the HH valence band for the three doped samples. Note that hole ISB transitions 1→2 and 2→3 are represented by downward arrows. (d) Measured SHG power vs. pump power density at focus.*

The experimental setup is sketched in Fig. 4(a) (see Methods for details). We estimate a peak power density at focus for the non-attenuated pump $I_{QC,max}$ = 23 W/cm$^2$ (see Supplementary Information S4). The resulting emitted SHG power $P_{SHG}$ is plotted in Fig. 4(d) as a function of $I_{QCL}$: a square-law relation is fitted to the integrated power data $P_{SHG} = K_2 P_{QCL}^2$ (see Supplementary Information S5). The value of $K_2$ in W$^{-1}$ reported in Table 1 is a measure of the SHG generation efficiency, as it is proportional to the square of $\chi^{(2)}_{zzz}(\nu)$ calculated at $\nu_{pump}$ = 29.3 THz. The resonance of $2\nu_{pump}$ with 1→3 of sample A then explains its highest $K_2$. Sample B is also resonant at 1→3, but its doping level is almost half of that of sample A and therefore its $K_2$ is approximately four times lower than that of sample A. Finally, sample C, in spite of the same doping level as sample A, has the lowest $K_2$, apparently because it is not doubly resonant. This interpretation is supported by calculations based on Eq. (1) (see Supplementary Information S4). We note that a maximum dimensionless conversion efficiency $P_{SHG} / P_{QCL}$ = 1.5×10$^{-4}$ at $I_{QCL,max}$ = 23 W/cm$^2$ in sample A is comparable to that found in recent experiments performed with a similar QCL pump on InGaAs/AlInAs ACQWs with comparable number of periods [*23, 24*].

**SHG emission at room temperature**

In order to clearly demonstrate nonlinear emission at room temperature, we pumped the samples with very high peak-power pulses from a nonlinear optical parametric amplifier (NOPA) driving a difference-frequency generation (DFG) setup in Fig. 5(a) [*45*]. The maximum power density at focus inside the sample $I_{DFG}$ = 9×10$^7$ W/cm$^2$ is more than six orders of magnitude higher than that of the QCL pump. The tunability and the high power of this optical pump permit the investigation of SHG efficiency in the entire frequency range where Eq. 1 provides significant values, and not only at the peak value as for the QCL pump. In particular, we used the three DFG pulses in Fig.

5(b-c) with center frequency $\nu_{DFG,1}$=25 THz ($\nu_{pump}$ slightly off resonance), $\nu_{DFG,2}$=29 THz ($2\nu_{pump}$ at resonance with $\nu_{13}$) and $\nu_{DFG,3}$=32 THz ($\nu_{pump}$ at resonance with $\nu_{12}$). In waveguides containing ACQWs, when pumping at resonance with $\nu_{12}$, the pump intensity is rapidly depleted, ending up in an interaction region shorter than the waveguide length. Therefore, in ACQWs it is generally convenient to pump in exact resonance with $\nu_{13}/2$, instead. This scenario is approximately confirmed by the SHG spectra for sample A: the red dashed line in Fig. 5(d) represents the $\chi^{(2)}_{zzz}$ calculation based on Eq. 1 (not accounting for pump depletion) using the data from Fig. 3. This calculation predicts approximately equal SHG efficiencies for $\nu_{DFG,1}$ and $\nu_{DFG,3}$, however the SHG efficiency observed in Fig. 5(d) with the pump tuned at $\nu_{DFG,3}$ is three times lower than at $\nu_{DFG,1}$. The shaded grey Lorentzian curve represents the effect of the 1→2 ISBT in depleting the pump beam, which is obviously stronger at $\nu_{DFG,3}$; one can also see that even at $\nu_{DFG,2}$ the ideal SHG efficiency is not reached due to pump depletion. We note that, at these very high pump intensities, 1→2 absorption saturation and free carrier heating (*i.e.* high-$k_{//}$ states occupied by holes in the subband 1, leading to broader effective transition linewidths) can also play a role in decreasing the observed SHG efficiency from the ideal value.

Finally, in Fig. 5(e) the relative efficiency of samples A, B and C when pumping at $\nu_{DFG,2} \approx \nu_{QCL}$ compares very well with that of Fig. 4(c-d): sample B is less efficient than sample A due to lower doping level, sample C because it is not doubly-resonant. No SHG emission was observed from the undoped sample D.

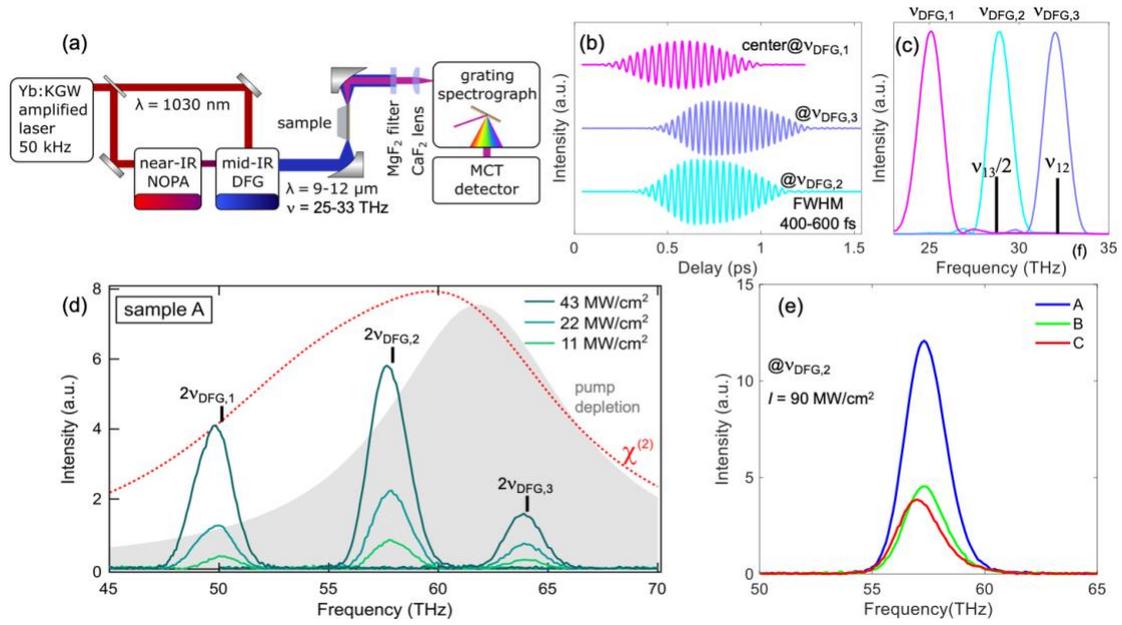

*Figure 5. SHG performance with high-intensity ultrashort pulses. (a) Experimental setup based on difference-frequency generation (DFG) and spectrally resolved detection with a mercury-cadmium telluride (MCT) diode. (b,c) Field-time profiles obtained by electro-optic sampling and corresponding FT spectrum of MIR pump pulses employed in the experiment. (d) Continuous curves: SHG spectra measured on sample A at different pump frequency and intensity. Red dashed curve: calculation of the nonlinear susceptibility according to Eq. 1. Grey shaded area: Absorption spectrum of the first ISBT, which results in depletion of the pump beam power. (e) SHG efficiency of different samples at optimal pumping conditions (i.e. $\nu_{DFG,2}$ and highest pulse power). All curves in this figure have arbitrary units (a.u.), i.e. only their spectral shape has a physical significance.*

**Absolute SHG efficiency**

In Fig. 6 we plot, for sample A, the emitted SHG peak power vs. the pump power density for both QCL and DFG pumps, in a log-log scale. It can be seen that the slope of the QCL data, corresponding to a quadratic dependence on the input pump power density, is not maintained in the DFG pump power range. An almost linear dependence is observed instead, a fact that we relate to a concurring two-photon absorption (TPA) effect that, at very high pump intensities, becomes a non-negligible competitor of the SHG process [5]. In quantum well systems displaying ISBTs, the radiative decay of the excited states (photoluminescence) is negligible if compared to the non-radiative decay, therefore the emission spectrum shows only the SHG line. The competition between TPA and SHG sets an ultimate high pump-power limit to SHG efficiency in SiGe ACQWs.

The experimental value of $\chi^{(2)}$ to be compared with calculations should then be extracted from the QCL pump experiment using the definition of $\chi^{(2)}=F_{2\omega}/F_\omega^2$, with $F$ electric field intensity at focus. Using $K_2$ of sample A from Table 1, and considering 20 periods of ACQWs, we find $\chi^{(2),exp}=5\times10^4$ pm/V (see Supplementary Information S4). This value compares well with the predicted peak value at 10 K of $\chi^{(2),theo}=3\times10^4$ pm/V (see Fig. 1(e)), the disagreement being probably related to uncertainty in the absolute detector signal calibration.

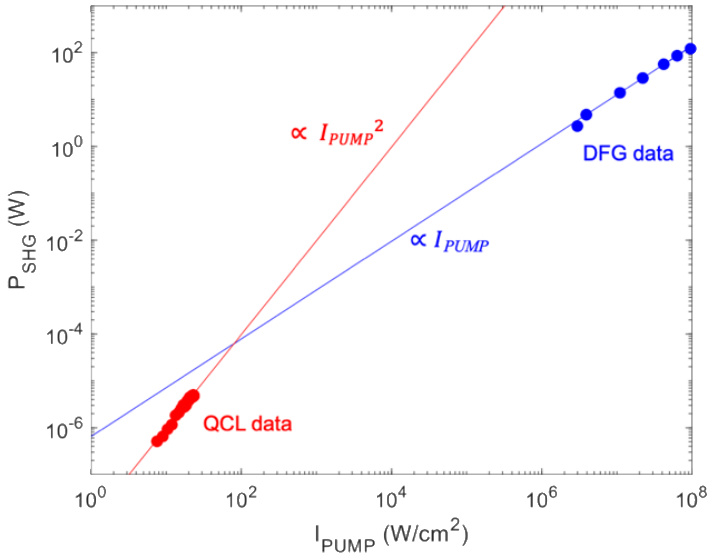

*Figure 6.* *Log-log plot of SHG emission power vs. pump power density for all QCL and DFG pump experiments performed on sample A. A quadratic slope has been superimposed to the QCL data, and a linear slope to the DFG data.*

A final consideration is worth on phase matching. In nonlinear crystals, the difference in the phase velocity $c/n(\nu)$ between pump and second-harmonic defines a maximum length along the waveguide over which the SHG power adds up coherently to that generated in previous positions along the waveguide, thereby imposing an upper limit to the SHG efficiency. In III-V compound semiconductor ACQWs at mid-IR frequencies, the dispersion due to the strong polar optical phonon absorption at around 10 THz limits SHG phase coherence. Instead, in SiGe ACQWs, which feature a nonpolar crystal lattice, the dispersion due to phonons is totally absent. In integrated waveguides, the modes will still have a dispersion due to the ratio between λ and the geometrical parameters of the waveguides, but in the presently employed slab waveguide this source of dispersion is negligible. There also exists a residual dispersion generated by the ISBT absorption resonances themselves, but this phenomenon can be neglected in the case of detuning from exact resonances, as for example when pumping the sample A at 25 and 29 THz. Therefore, in most of the present experiments homogeneous phase matching conditions apply, without having to include specific dispersion-compensation structures.

**Conclusions**

We have designed, grown and optically analyzed hole-doped Ge/SiGe asymmetric-coupled quantum wells optimized for second harmonic generation in silicon photonics chips. A giant second-order nonlinearity, four orders of magnitude higher than that of any natural nonlinear material, is observed with laser pump wavelengths in the 9.2 to 12 µm range (second harmonic in the 4.6 to 6 µm range), well within the even broader transparency range of existing Ge-rich integrated waveguides that have been developed for on-chip spectroscopic sensing applications. The giant nonlinearity is attributed to the double-resonance effect of the laser pump photon energy and the SHG photon energy with the intersubband transitions among three levels equally spaced in energy. The absolute nonlinear susceptibility has been determined at low temperatures and found to be in good agreement with wavefunction calculations. Room-temperature operation with ultrashort pulses has shown that the optimal SHG efficiency is obtained for small red-detuning from the double resonance. The material growth technology and the infrared wavelength range employed here are compatible with the standard silicon photonics foundry processes, opening the way to the on-chip integration of second-order optical nonlinearities, to be exploited in future molecular sensing and free-space telecommunications devices.

**Methods**

*Theoretical calculations*

The valence band structure has been calculated relying on a semiempirical tight-binding Hamiltonian model in the first neighbour approximation [46, 47]. The adopted basis set includes $sp^3d^5s^*$ orbitals in both spin configurations. Self and hopping energies together with scaling exponents, used to account for the strain induced lattice distortion are reported in [48] and [49] for Si and Ge, respectively. In order to properly describe non-parabolicity effects, the linear absorption and second harmonic susceptivity spectra have been calculated performing a 3D sampling of the BZ zone in a neighbour of the Γ point. Dipole matrix elements have been estimated following the procedure given in [46]. $\chi^{(2)}_{lmn}(2\nu)$ has been calculated assuming that holes populate the fundamental subband only following Eq. 1 of Ref. [25] which describes both resonant and non-resonant contributions. A lorentzian lineshape with a HWHM of 5 meV has been used to phenomenologically describe the line broadening of the ISBTs.

*Sample growth and structural characterization*

Samples A-D were grown by low energy plasma enhanced chemical vapor deposition (LEPECVD) [42] on a 100 mm intrinsic Si(001) substrates with a resistivity > 6000 Ω cm. Before heteroepitaxy, the native oxide was removed by dipping the substrate in aqueous HF solution (HF:H$_2$O 1:10) for 30 s. The first part of the structure consists of a 1 μm thick Si$_{0.8}$Ge$_{0.2}$ layer, deposited at a rate of 5 nm/s at a temperature of 600°C. The deposition was followed by an in-situ thermal annealing at 700°C for 30 minutes in order to promote the full relaxation of the layer. The second part of the structure consists of a 1 μm thick Si$_{0.3}$Ge$_{0.7}$ deposited at 5 nm/s at 500°C forming a fully relaxed virtual substrate (VS) for the ACQW stack. The ACQW structure has been grown at a rate ~ 0.1 nm/s at a temperature of 350°C to minimize the intermixing, and it consists of 20 repetitions of the following sequence: Si$_{0.1}$Ge$_{0.9}$ main well/Si$_{0.4}$Ge$_{0.6}$ tunneling barrier/Si$_{0.25}$Ge$_{0.75}$ secondary well/Si$_{0.4}$Ge$_{0.6}$ main barrier. The main well has been p-doped in-situ by adding B$_2$H$_6$ during the growth. The thickness of the main barrier has been designed to obtain a mean Ge concentration in the ACQW stack equal to the one of the VS. In this way the compressive strain of the wells is compensated by the tensile strain in the barriers, thus obtaining a strain-symmetrized structure.

A XRD apparatus was used to record the (224) and (004) reciprocal space maps of sample A are shown in Fig. 2(a) and 2(b), respectively. Fig. 2(c) shows the (004) ω–2θ scan of Sample A, together with multi-beam dynamical Darwin model simulations [50, 51] as implemented in the software "xrayutilities" [52]. This software has been used to retrieve the compositional profile of Fig. 2(f). A different piece of sample A has been further characterized by STEM and EDS. A JEOL Monochoromated ARM200F TEM equipped with a monochromator, probe and image aberration correctors, and double silicon drift detector. EDS detectors was operated at 200 kV for STEM annular dark field (ADF) imaging and EDS analysis. The beam convergence semi-angle used for STEM and EDS analysis was ~ 27 mrad. The inner collection semi-angle for ADF imaging was set to ~ 55 mrad. The electron beam size was estimated to ~1 Å in diameter. The STEM image (Fig 2(c)) shows ACQWs with sharp interfaces arranged in identical subsequent periods along the growth direction $z$.

*IR spectroscopy*

A FTIR spectrometer (Bruker Vertex 70v) was used to collect the TM and TE transmission spectra, equipped with a wideband beamsplitter (Bruker), a broad linearity range MCT detector (Bruker), a liquid-He flow optical cryostat (Janis Research co.) and a wire-grid lithographic KRS5 polarizer (Bruker). The transmitted spectral intensity is measured for both TM and TE polarizations (electric field parallel and orthogonal to $z$, respectively). The dichroic absorption is calculated as $A = -\ln(TM/TE)$, therefore peaks correspond to $z$-dipole ISBTs and dips to $xy$-dipole ISBTs. The narrow $\gamma_{12} < 2$ THz obtained at T = 10 K for all samples confirms the high quality of the heterostructures. At T = 10 K, only the fundamental subband $E_1$ is populated, so the integrated spectral weight $\sum_i f_{1i}$ provides an estimate of the total hole sheet-density [33] in reasonable agreement with the nominal doping values (one has $N_1 \approx N_a$ see Table 1). The spectral weight ratios $f_{13}/f_{12} = 0.96, 0.82$ for samples A, B respectively, are reasonably close to unity, which represents the optimal value for SHG in ACQWs as explained in [53], while we get $f_{13}/f_{12} = 1.33$ for the purposely detuned sample C.

For the first SHG experiment at 10 K, a monochromatic distributed feedback-grating QCL emitting at λ = 10.3 μm, or $\nu_{QCL}$=29.3 THz ≈ $\nu_{13}$/2 was used to pump the SHG emission at T = 10 K, where ISBT linewidths are narrowest. ZnSe lenses and wire-grid KRS5 polarizers were employed, together with the same optical cryostat of the FTIR setup. A thermoelectric-cooled MCT detector with bandgap at λ = 8 μm was used to intrinsically reject the signal due to the pump photons, but thermal excitation of the detector by the pump beam could not be avoided. The resulting linear background was subtracted by offsetting the sample position along the optical axis (z-scan technique). The MCT detector and the subsequent lock-in amplification chain was calibrated with a second QCL emitting at λ = 5.7 μm. The pump electric field was vertically polarized (TM), so only the tensor component $\chi^{(2)}_{zzz}$ was probed. The pump power was progressively attenuated from its maximum peak power $P_{QCL}$ = 30 mW without varying the polarization direction, using the cross-polarizer technique shown in Fig. 4a (see Supplementary Information S5). The collimated QCL beam was focused into the cryostat precisely at the center of the slab waveguide, so as to maximize the radiation electric field $F_z$ in the ACQW layer.

For the second SHG experiment at 300 K, strong multi-cycle pump pulses were generated as described in [45], centered at $\nu_{DFG,1}$ = 25 THz, $\nu_{DFG,2}$ = 29 THz or $\nu_{DFG,3}$ = 32 THz, and featuring a Fourier-transform-limited bandwidth $\Delta\nu_{DFG}$ ≈ 2 THz. The radiation exiting the waveguide was spectrally resolved with a mid-IR grating spectrometer in the range of 50-75 THz. Because $\Delta\nu_{DFG}$ values are narrower than the ISBT linewidths at 300 K, they define the output SHG spectrum bandwidths equal to 2$\Delta\nu_{DFG}$. Due to the high peak power density at focus up to 9×10$^7$ W/cm$^2$, the hot carrier temperature after optical pump absorption is expected to surpass the lattice temperature at 300 K hence making cryogenic cooling unessential.

All power values are estimated inside a generic sample, taking into account the reflection losses of 0.31 at the 20° Si-prism facet and, for the QCL experiment, the reflections at the KRS5 cryostat windows and polarizers.

**Acknowledgements and funding**

This project has received funding from Sapienza University of Rome, Ricerca d'Ateneo 2017. This project has received funding from the European Union's Horizon 2020 research and innovation program under grant agreements No.823717-ESTEEM3 and No.766955-microSPIRE.

**Author contributions**

J.F., M.V. and M.O. conceived the experiment. J.F., A.B. and V.F. performed the epitaxial growth. D.C. performed the XRD characterization. L.Z and E.O. performed the TEM characterization. M.O., C.C., A.M. and L.B. conducted the FTIR measurements and the QCL experiment. J.K., J.A. and D.B. conducted the DFG experiment. M.V. and M. D. S. designed the structures and performed the simulations. M.O, M.V., C.C. and J.F. wrote the paper with input from all authors.

**Competing interests**

The authors declare no competing interests.

**ORCID for corresponding authors**

Corresponding authors: Michele Ortolani, ORCID: 0000-0002-7203-5355 and Michele Virgilio, ORCID: 0000-0002-7847-6813

**Figure legends:**

**Figure 1**: Epitaxial structure of the sample, band-structure, in-plane dispersion of the relevant wavefunction and calculated $\chi^{(2)}_{zzz}$.

**Figure 2**: Material structural characterization. Reciprocal space maps and (004) ω-2θ scan measured by HR-XRD. STEM images. Reconstructed compositional profile.

**Figure 3**: Linear dichroic absorption spectroscopy in the mid-IR. Dichroic absorption as measured by FTIR for all the samples at 10 K and at room T. Comparison between the experimental and simulated dichroic absorption for sample A at 10 K and 300 K.

**Figure 4**: absolute characterization of SHG efficiency. Experimental set-up and plot of the second harmonic power as a function of the input intensity.

**Figure 5**: SHG with high-intensity ultrashort pulses. Experimental set-up. Field-time profiles and corresponding FT spectrum of pump pulses. SHG spectra measured with sample A at different pump frequency and intensity. SHG efficiency of different samples at optimal pumping conditions.

**Figure 6**: plot of SHG emission power vs. pump power density for all QCL and DFG pump experiments performed on sample A.

**Tables**

**Table 1**. Summary of design and measured parameters of the four samples

| Sample | Wavefunction design | Main well thickness $t_w$ (nm) | Acceptor density $N_a$ (cm$^{-2}$) | Measured hole density $N_{tot}$ (cm$^{-2}$) | Measured SHG efficiency $K_2$ (1/W) |
|---|---|---|---|---|---|
| A | Doubly resonant | 2.7 | 4.2 ×10$^{11}$ | (3.9±0.3) ×10$^{11}$ | 5.7×10$^{-3}$ |
| B | Doubly resonant | 2.7 | 2.1×10$^{11}$ | (2.5±0.3) ×10$^{11}$ | 1.4×10$^{-3}$ |
| C | Single resonance | 3.0 | 4.2 ×10$^{11}$ | (3.9±0.3) ×10$^{11}$ | 0.6×10$^{-3}$ |
| D | Reference | 2.7 | undoped | - | < 0.1×10$^{-3}$ |